\let\includefigures=\iftrue
\let\useblackboard=\iftrue
\newfam\black
\input harvmac

\noblackbox

%Figure Stuff
\includefigures
\message{If you do not have epsf.tex (to include figures),}
\message{change the option at the top of the tex file.}
\input epsf
\def\figin{\epsfcheck\figin}\def\figins{\epsfcheck\figins}
\def\epsfcheck{\ifx\epsfbox\UnDeFiNeD
\message{(NO epsf.tex, FIGURES WILL BE IGNORED)}
\gdef\figin##1{\vskip2in}\gdef\figins##1{\hskip.5in}% blank space instead
\else\message{(FIGURES WILL BE INCLUDED)}%
\gdef\figin##1{##1}\gdef\figins##1{##1}\fi}
\def\DefWarn#1{}
\def\figinsert{\goodbreak\midinsert}
\def\ifig#1#2#3{\DefWarn#1\xdef#1{fig.~\the\figno}
\writedef{#1\leftbracket fig.\noexpand~\the\figno}%
\figinsert\figin{\centerline{#3}}\medskip\centerline{\vbox{
\baselineskip12pt\advance\hsize by -1truein
\noindent\footnotefont{\bf Fig.~\the\figno:} #2}}
\bigskip\endinsert\global\advance\figno by1}
%%%
\else
\def\ifig#1#2#3{\xdef#1{fig.~\the\figno}
\writedef{#1\leftbracket fig.\noexpand~\the\figno}%
%\figinsert\figin{\centerline{#3}}\medskip
%\centerline{\vbox{\baselineskip12pt
%\advance\hsize by -1truein\noindent
%\footnotefont{\bf Fig.~\the\figno:} #2}}
%\bigskip\endinsert
\global\advance\figno by1} \fi
%

%%BLACKBOARD FONT STUFF
\useblackboard
\message{If you do not have msbm (blackboard bold) fonts,}
\message{change the option at the top of the tex file.}
\font\blackboard=msbm10 scaled \magstep1 \font\blackboards=msbm7
\font\blackboardss=msbm5 \textfont\black=\blackboard
\scriptfont\black=\blackboards
\scriptscriptfont\black=\blackboardss

\else

\fi
% *************************************

%
\def\yboxit#1#2{\vbox{\hrule height #1 \hbox{\vrule width #1
\vbox{#2}\vrule width #1 }\hrule height #1 }}
\def\fillbox#1{\hbox to #1{\vbox to #1{\vfil}\hfil}}
\def\ybox{{\lower 1.3pt \yboxit{0.4pt}{\fillbox{8pt}}\hskip-0.2pt}}

\noblackbox

%Figure Stuff
\includefigures
\message{If you do not have epsf.tex (to include figures),}
\message{change the option at the top of the tex file.}
\input epsf
\def\figin{\epsfcheck\figin}\def\figins{\epsfcheck\figins}
\def\epsfcheck{\ifx\epsfbox\UnDeFiNeD
\message{(NO epsf.tex, FIGURES WILL BE IGNORED)}
\gdef\figin##1{\vskip2in}\gdef\figins##1{\hskip.5in}% blank space instead
\else\message{(FIGURES WILL BE INCLUDED)}%
\gdef\figin##1{##1}\gdef\figins##1{##1}\fi}
\def\DefWarn#1{}
\def\figinsert{\goodbreak\midinsert}
\def\ifig#1#2#3{\DefWarn#1\xdef#1{fig. \the\figno}
\writedef{#1\leftbracket fig.\noexpand \the\figno}%
\figinsert\figin{\centerline{#3}}\medskip\centerline{\vbox{
\baselineskip12pt\advance\hsize by -1truein
\noindent\footnotefont{\bf Fig. \the\figno:} #2}}
\bigskip\endinsert\global\advance\figno by1}
%%%
\else
\def\ifig#1#2#3{\xdef#1{fig. \the\figno}
\writedef{#1\leftbracket fig.\noexpand \the\figno}%
%\figinsert\figin{\centerline{#3}}\medskip
%\centerline{\vbox{\baselineskip12pt
%\advance\hsize by -1truein\noindent
%\footnotefont{\bf Fig. \the\figno:} #2}}
%\bigskip\endinsert
\global\advance\figno by1} \fi
%

%%BLACKBOARD FONT STUFF
\useblackboard
\message{If you do not have msbm (blackboard bold) fonts,}
\message{change the option at the top of the tex file.}
\font\blackboard=msbm10 scaled \magstep1 \font\blackboards=msbm7
\font\blackboardss=msbm5 \textfont\black=\blackboard
\scriptfont\black=\blackboards
\scriptscriptfont\black=\blackboardss

\else

\fi
% *************************************
%\draft
%
\def\yboxit#1#2{\vbox{\hrule height #1 \hbox{\vrule width #1
\vbox{#2}\vrule width #1 }\hrule height #1 }}
\def\fillbox#1{\hbox to #1{\vbox to #1{\vfil}\hfil}}
\def\ybox{{\lower 1.3pt \yboxit{0.4pt}{\fillbox{8pt}}\hskip-0.2pt}}

\def\d{\delta}
\def\m{\mu}

\def\m{\mu}

\def\G{\Gamma}
\def\g{\gamma}
\def\a{\alpha}

\def\ee{\eqn\placeholder }

\def\b{\beta}

\def\IZ{\relax\ifmmode\mathchoice
{\hbox{\cmss Z\kern-.4em Z}}{\hbox{\cmss Z\kern-.4em Z}}
{\lower.9pt\hbox{\cmsss Z\kern-.4em Z}} {\lower1.2pt\hbox{\cmsss
Z\kern-.4em Z}}\else{\cmss Z\kern-.4em Z}\fi} \font\cmss=cmss10
\font\cmsss=cmss10 at 7pt
\def\inbar{\,\vrule height1.5ex width.4pt depth0pt}
\def\IC{{\relax\hbox{$\inbar\kern-.3em{\rm C}$}}}
\def\IQ{{\relax\hbox{$\inbar\kern-.3em{\rm Q}$}}}
\def\IP{\relax{\rm I\kern-.18em P}}

%%%%%%%EXTRA STUFF%%%%%%%%%
\def\IR{{\bf R}}
\def\CP{{\bf CP}}
%Other

\def\ap{\alpha'}

\def\CM{{\cal M}}
\def\CW{{\cal W}}
\def\CN{{\cal N}}
\def\p{{\partial}}
\def\CS{{\cal S}}
\def\rmx{{\rm x}}

\def\irc{{}^{\rm{\underline{c}}}}
%%%%%%%%%%%%%%%%%%%%%%%%%%%%%%
%0000000000
%\draft

%%%%%%%%%%%%%%%%%%%%%%%%
%rrrrrrrrrr

%BEGINREFERENCES

%\AspinwallMN
\lref\AspinwallMN{ P.~S.~Aspinwall,
``K3 surfaces and string
duality,'' arXiv:hep-th/9611137.
%%CITATION = HEP-TH 9611137;%%
}

%\GreeneCY
\lref\GreeneCY{ B.~R.~Greene, ``String theory on Calabi-Yau
manifolds,'' arXiv:hep-th/9702155.
%%CITATION = HEP-TH 9702155;%%
}

%\Polchinski
\lref\Polchinski{ J.~Polchinski, ``String Theory,'' {\it
Cambridge, UK: Univ. Pr. (1998)}. }

%\AspinwallFD
\lref\AspinwallFD{ P.~S.~Aspinwall, ``Compactification, geometry
and duality: N = 2,'' arXiv:hep-th/0001001.
%%CITATION = HEP-TH 0001001;%%
}

%\LiuFT
\lref\LiuFT{ H.~Liu, G.~Moore and N.~Seiberg, ``Strings in a
time-dependent orbifold,'' JHEP {\bf 0206}, 045 (2002)
[arXiv:hep-th/0204168].
%%CITATION = HEP-TH 0204168;%%
}

%\HorowitzAP
\lref\HorowitzAP{ G.~T.~Horowitz and A.~R.~Steif, ``Singular
String Solutions With Nonsingular Initial Data,'' Phys.\ Lett.\ B
{\bf 258}, 91 (1991).
%%CITATION = PHLTA,B258,91;%%
}

%\Figueroa
\lref\Figueroa{ J.~Figueroa-O'Farrill and J.~Simon, ``Generalized
supersymmetric fluxbranes,'' JHEP {\bf 0112}, 011 (2001)
[arXiv:hep-th/0110170].
%%CITATION = HEP-TH 0110170;%%
}

%\SimonMA
\lref\SimonMA{ J.~Simon, ``The geometry of null rotation
identifications,'' JHEP {\bf 0206}, 001 (2002)
[arXiv:hep-th/0203201].
%%CITATION = HEP-TH 0203201;%%
}

%\LawrenceAJ
\lref\LawrenceAJ{ A.~Lawrence, ``On the instability of 3D null
singularities,'' arXiv:hep-th/0205288.
%%CITATION = HEP-TH 0205288;%%
}

%\LiuKB
\lref\LiuKB{ H.~Liu, G.~Moore and N.~Seiberg, ``Strings in
time-dependent orbifolds,'' JHEP {\bf 0210}, 031 (2002)
[arXiv:hep-th/0206182].
%%CITATION = HEP-TH 0206182;%%
}

%\FabingerKR
\lref\FabingerKR{ M.~Fabinger and J.~M$\irc$Greevy, ``On smooth
time-dependent orbifolds and null singularities,''
arXiv:hep-th/0206196.
%%CITATION = HEP-TH 0206196;%%
}

%\HorowitzMW
\lref\HorowitzMW{ G.~T.~Horowitz and J.~Polchinski, ``Instability
of spacelike and null orbifold singularities,''
arXiv:hep-th/0206228.
%%CITATION = HEP-TH 0206228;%%
}

%\GreeneTX
\lref\GreeneTX{ B.~R.~Greene and Y.~Kanter, ``Small volumes in
compactified string theory,'' Nucl.\ Phys.\ B {\bf 497}, 127
(1997) [arXiv:hep-th/9612181].
%%CITATION = HEP-TH 9612181;%%
}

%\PolchinskiSM
\lref\PolchinskiSM{ J.~Polchinski and A.~Strominger, ``New Vacua
for Type II String Theory,'' Phys.\ Lett.\ B {\bf 388}, 736 (1996)
[arXiv:hep-th/9510227].
%%CITATION = HEP-TH 9510227;%%
}

%\CandelasRM
\lref\CandelasRM{ P.~Candelas, X.~C.~De La Ossa, P.~S.~Green and
L.~Parkes, ``A Pair Of Calabi-Yau Manifolds As An Exactly Soluble
Superconformal  Theory,'' Nucl.\ Phys.\ B {\bf 359}, 21 (1991);
``An Exactly Soluble Superconformal Theory From A Mirror Pair Of
Calabi-Yau Manifolds,'' Phys.\ Lett.\ B {\bf 258}, 118 (1991).
%%CITATION = NUPHA,B359,21;%%
}

%%\CandelasQD
\lref\CandelasQD{
P.~Candelas, X.~C.~De la Ossa, P.~S.~Green and L.~Parkes,
``An Exactly Soluble Superconformal Theory From A Mirror Pair Of 
Calabi-Yau Manifolds,''
Phys.\ Lett.\ B {\bf 258}, 118 (1991).
%%%CITATION = PHLTA,B258,118;%%
}

%\GreenSP
\lref\GreenSP{
M.~B.~Green, J.~H.~Schwarz and E.~Witten,
%``Superstring Theory. Vol. 1: Introduction,''
{\it  Cambridge, Uk: Univ. Pr. ( 1987) 469 P. } }

%%%%%%%%%%%%%%%%%%%%%%
%More references
%zzzzzzzzzz

%\KhouryBZ
\lref\KhouryBZ{ J.~Khoury, B.~A.~Ovrut, N.~Seiberg,
P.~J.~Steinhardt and N.~Turok, ``From big crunch to big bang,''
arXiv:hep-th/0108187;
%%CITATION = HEP-TH 0108187;%%
%\SeibergHR
N.~Seiberg, ``From big crunch to big bang - is it possible?,''
arXiv:hep-th/0201039.
%%CITATION = HEP-TH 0201039;%%
}

%\BalasubramanianRY
\lref\BalasubramanianRY{ V.~Balasubramanian, S.~F.~Hassan,
E.~Keski-Vakkuri and A.~Naqvi, ``A space-time orbifold: A toy
model for a cosmological singularity,'' arXiv:hep-th/0202187.
%%CITATION = HEP-TH 0202187;%%
}
%%\GutperleAI
%\lref\GutperleAI{ M.~Gutperle and A.~Strominger, ``Spacelike
%branes,'' arXiv:hep-th/0202210.
%%CITATION = HEP-TH 0202210;%%
%}
%\CornalbaFI
\lref\CornalbaFI{ L.~Cornalba and M.~S.~Costa, ``A New
Cosmological Scenario in String Theory,'' arXiv:hep-th/0203031.
%%CITATION = HEP-TH 0203031;%%
}
%\NekrasovKF
\lref\NekrasovKF{ N.~A.~Nekrasov, ``Milne universe, tachyons, and
quantum group,'' hep-th/0203112.
%%CITATION = HEP-TH 0203112;%%
}

%\lref\nullbranes{
%\Figueroa
\lref\Figueroa{ J.~Figueroa-O'Farrill and J.~Sim\'{o}n,
``Generalized supersymmetric fluxbranes,'' JHEP {\bf 0112}, 011
(2001) [arXiv:hep-th/0110170]}
%%CITATION = HEP-TH 0110170;%%
%
%\SimonMA
\lref\SimonMA{ J.~Sim\'{o}n, ``The geometry of null rotation
identifications,'' arXiv:hep-th/0203201.}
%%CITATION = HEP-TH 0203201;%%

%\CornalbaNV
\lref\CornalbaNV{ L.~Cornalba, M.~S.~Costa and C.~Kounnas, ``A
resolution of the cosmological singularity with orientifolds,''
arXiv:hep-th/0204261.
%%CITATION = HEP-TH 0204261;%%
}

%OLD STUFF
%\lref\otherbackgrounds{
%
%\AmatiSA
\lref\AmatiSA{ D.~Amati and C.~Klim\v{c}\'{\i}k, ``Nonperturbative
Computation Of The Weyl Anomaly For A Class Of Nontrivial
Backgrounds,'' Phys.\ Lett.\ B {\bf 219}, 443 (1989).}
%%CITATION = PHLTA,B219,443;%%
%
%\HorowitzBV
\lref\HorowitzBV{ G.~T.~Horowitz and A.~R.~Steif, ``Space-Time
Singularities In String Theory,'' Phys.\ Rev.\ Lett.\  {\bf 64},
260 (1990); ``Strings In Strong Gravitational Fields,'' Phys.\
Rev.\ D {\bf 42}, 1950 (1990).}
%%CITATION = PRLTA,64,260;%%
%

%\deVegaNR
\lref\deVegaNR{ H.~J.~de Vega and N.~Sanchez, ``Space-Time
Singularities In String Theory And String Propagation Through
Gravitational Shock Waves,'' Phys.\ Rev.\ Lett.\  {\bf 65}, 1517
(1990).}
%%CITATION = PRLTA,65,1517;%%
%
%\NappiKV
\lref\NappiKV{ C.~R.~Nappi and E.~Witten, ``A Closed, expanding
universe in string theory,'' Phys.\ Lett.\ B {\bf 293}, 309 (1992)
[arXiv:hep-th/9206078]; ``A WZW model based on a nonsemisimple
group,'' Phys.\ Rev.\ Lett.\  {\bf 71}, 3751 (1993)
[arXiv:hep-th/9310112].}
%%CITATION = HEP-TH 9206078;%%
%
%\NappiIE
%\lref\NappiIE{ C.~R.~Nappi and E.~Witten, ``A WZW model based on a
%nonsemisimple group,'' Phys.\ Rev.\ Lett.\  {\bf 71}, 3751 (1993)
%[arXiv:hep-th/9310112].}
%%CITATION = HEP-TH 9310112;%%
%
\lref\KiritsisJK{ E.~Kiritsis and C.~Kounnas, ``String Propagation
In Gravitational Wave Backgrounds,'' Phys.\ Lett.\ B {\bf 320},
264 (1994) [arXiv:hep-th/9310202]; ``Dynamical topology change,
compactification and waves in a stringy early universe,''
arXiv:hep-th/9407005.}
%%CITATION = HEP-TH 9310202;%%
%
%\KiritsisIJ
\lref\KiritsisIJ{ E.~Kiritsis, C.~Kounnas and D.~L\"ust,
``Superstring gravitational wave backgrounds with space-time
supersymmetry,'' Phys.\ Lett.\ B {\bf 331}, 321 (1994)
[arXiv:hep-th/9404114].}
%%CITATION = HEP-TH 9404114;%%

%END OF OLD STUFF

%
%\lref\newexamples{
%\lref\SilversteinXN{
%%\SilversteinXN
%E.~Silverstein, ``(A)dS backgrounds from asymmetric
%orientifolds,'' \ [arXiv:hep-th/0106209].}
%%CITATION = HEP-TH 0106209;%%
%
%\KoganNN
\lref\KoganNN{ I.~I.~Kogan and N.~B.~Reis, ``H-branes and chiral
strings,''
%Int.\ J.\ Mod.\ Phys.\ A {\bf 16}, 4567 (2001)
[arXiv:hep-th/0107163].}
%%CITATION = HEP-TH 0107163;%%
%

%\GutperleAI
\lref\GutperleAI{ M.~Gutperle and A.~Strominger, ``Spacelike
branes,'' [arXiv:hep-th/0202210].}
%%CITATION = HEP-TH 0202210;%%
%
%\BuchelWF
\lref\BuchelWF{ A.~Buchel, ``Gauge / gravity correspondence in
accelerating universe,'' arXiv:hep-th/0203041.}
%%CITATION = HEP-TH 0203041;%%

\lref\aharony{ O.~Aharony, M.~Fabinger, G.~Horowitz, and
E.~Silverstein, ``Clean Time-Dependent String Backgrounds from
Bubble Baths,'' arXiv:hep-th/0204158.}

%\ElitzurRT
\lref\ElitzurRT{ S.~Elitzur, A.~Giveon, D.~Kutasov and
E.~Rabinovici, ``From big bang to big crunch and beyond,''
arXiv:hep-th/0204189.}
%%CITATION = HEP-TH 0204189;%%
%
%\CornalbaNV
\lref\CornalbaNV{ L.~Cornalba, M.~S.~Costa and C.~Kounnas, ``A
resolution of the cosmological singularity with orientifolds,''
arXiv:hep-th/0204261.}
%%CITATION = HEP-TH 0204261;%%
%
%\CrapsII
\lref\CrapsII{ B.~Craps, D.~Kutasov and G.~Rajesh, ``String
propagation in the presence of cosmological singularities,''
arXiv:hep-th/0205101.}
%%CITATION = HEP-TH 0205101;%%
%

%
%\KachruKX
\lref\KachruKX{ S.~Kachru and L.~M$\irc$Allister, ``Bouncing brane
cosmologies from warped string compactifications,''
arXiv:hep-th/0205209.}
%%CITATION = HEP-TH 0205209;%%
%

%\LawrenceAJ
\lref\LawrenceAJ{ A.~Lawrence, ``On the instability of 3D null
singularities,'' arXiv:hep-th/0205288.
%%CITATION = HEP-TH 0205288;%%
}

\lref\ppwaves{J.~Ehlers and W.~Kundt,
%``Exact Solutions of the Gravitational Field Equations,''
in ``Gravitation: an introduction to current research,'' ed.
L.~Witten (1962); See also Misner, Thorne and Wheeler,
``Gravitation,'' Ex. 35.8. }

\lref\rohm{R. Rohm, ``Spontaneous supersymmetry breaking in
supersymmetric string theories,''  Nucl.\ Phys.\ B {\bf 237}, 553
(1984). }

%\MartinecXQ
\lref\MartinecXQ{ E.~J.~Martinec and W.~M$\irc$Elgin, ``Exciting AdS
orbifolds,''
%JHEP {\bf 0210}, 050 (2002)
[arXiv:hep-th/0206175].
%%CITATION = HEP-TH 0206175;%%
}

%\CaiMR
\lref\CaiMR{ R.~G.~Cai, ``Constant curvature black hole and dual
field theory,'' Phys.\ Lett.\ B {\bf 544}, 176 (2002)
[arXiv:hep-th/0206223].
%%CITATION = HEP-TH 0206223;%%
}

%\MaldacenaFY
\lref\MaldacenaFY{ J.~Maldacena and L.~Maoz, ``Strings on pp-waves
and massive two dimensional field theories,''
arXiv:hep-th/0207284.
%%CITATION = HEP-TH 0207284;%%
}

%\BuchelKJ
\lref\BuchelKJ{ A.~Buchel, P.~Langfelder and J.~Walcher, ``On
time-dependent backgrounds in supergravity and string theory,''
arXiv:hep-th/0207214.
%%CITATION = HEP-TH 0207214;%%
}

%\HemmingKD
\lref\HemmingKD{ S.~Hemming, E.~Keski-Vakkuri and P.~Kraus,
``Strings in the extended BTZ spacetime,'' JHEP {\bf 0210}, 006
(2002) [arXiv:hep-th/0208003].
%%CITATION = HEP-TH 0208003;%%
}

%\HashimotoNR
\lref\HashimotoNR{ A.~Hashimoto and S.~Sethi, ``Holography and
string dynamics in time-dependent backgrounds,''
arXiv:hep-th/0208126.
%%CITATION = HEP-TH 0208126;%%
}

%\SimonCF
\lref\SimonCF{ J.~Simon, ``Null orbifolds in AdS, time dependence
and holography,'' JHEP {\bf 0210}, 036 (2002)
[arXiv:hep-th/0208165].
%%CITATION = HEP-TH 0208165;%%
}

%\AlishahihaBK
\lref\AlishahihaBK{ M.~Alishahiha and S.~Parvizi, ``Branes in
time-dependent backgrounds and AdS/CFT correspondence,'' JHEP {\bf
0210}, 047 (2002) [arXiv:hep-th/0208187].
%%CITATION = HEP-TH 0208187;%%
}

%\SatohNJ
\lref\SatohNJ{ Y.~Satoh and J.~Troost, ``Massless BTZ black holes
in minisuperspace,'' JHEP {\bf 0211}, 042 (2002)
[arXiv:hep-th/0209195].
%%CITATION = HEP-TH 0209195;%%
}

%\DolanPX
\lref\DolanPX{ L.~Dolan and C.~R.~Nappi, ``Noncommutativity in a
time-dependent background,'' arXiv:hep-th/0210030.
%%CITATION = HEP-TH 0210030;%%
}

%\CaiSV
\lref\CaiSV{ R.~G.~Cai, J.~X.~Lu and N.~Ohta, ``NCOS and D-branes
in time-dependent backgrounds,'' Phys.\ Lett.\ B {\bf 551}, 178
(2003) [arXiv:hep-th/0210206].
%%CITATION = HEP-TH 0210206;%%
}

%\HubenyPJ
\lref\HubenyPJ{ V.~E.~Hubeny and M.~Rangamani, ``No horizons in
pp-waves,'' JHEP {\bf 0211}, 021 (2002) [arXiv:hep-th/0210234];
``Causal structures of pp-waves,'' arXiv:hep-th/0211195;
``Generating asymptotically plane wave spacetimes,''
arXiv:hep-th/0211206.
%%CITATION = HEP-TH 0210234;%%
}

%\BachasQT
\lref\BachasQT{ C.~Bachas and C.~Hull, ``Null brane
intersections,''
% JHEP {\bf 0212}, 035 (2002)
[arXiv:hep-th/0210269].
%%CITATION = HEP-TH 0210269;%%
}

%\MyersBK
\lref\MyersBK{ R.~C.~Myers and D.~J.~Winters, ``From D - anti-D
pairs to branes in motion,'' arXiv:hep-th/0211042.
%%CITATION = HEP-TH 0211042;%%
}

%\OkuyamaPC
\lref\OkuyamaPC{ K.~Okuyama, ``D-branes on the null-brane,''
arXiv:hep-th/0211218.
%%CITATION = HEP-TH 0211218;%%
}

%\AlishahihaBC
\lref\AlishahihaBC{ M.~Alishahiha, M.~M.~Sheikh-Jabbari and
R.~Tatar, ``Spacetime quotients, Penrose limits and conformal
symmetry restoration,'' arXiv:hep-th/0211285.
%%CITATION = HEP-TH 0211285;%%
}

%\PapadopoulosBG
\lref\PapadopoulosBG{ G.~Papadopoulos, J.~G.~Russo and
A.~A.~Tseytlin, ``Solvable model of strings in a time-dependent
plane-wave background,'' arXiv:hep-th/0211289.
%%CITATION = HEP-TH 0211289;%%
}

%\ChoGA
\lref\ChoGA{ J.~H.~Cho and P.~Oh, ``Supersymmetric boost on
intersecting D-branes,'' arXiv:hep-th/0212009.
%%CITATION = HEP-TH 0212009;%%
}

%\GauntlettFZ
\lref\GauntlettFZ{ J.~P.~Gauntlett and S.~Pakis, ``The geometry of
D = 11 Killing spinors,'' arXiv:hep-th/0212008.
%%CITATION = HEP-TH 0212008;%%
}

%\TseytlinVA
%\lref\TseytlinVA{
\lref\TseytlinABC{
 A.~A.~Tseytlin,
``A Class of finite two-dimensional sigma models and string
vacua,''
%Phys.\ Lett.\ B {\bf 288}, 279 (1992)
[arXiv:hep-th/9205058];
%%CITATION = HEP-TH 9205058;%%
%}
%\TseytlinEE
%\lref\TseytlinEE{
%A.~A.~Tseytlin,
 ``Finite sigma models and exact string solutions
with Minkowski signature metric,''
%Phys.\ Rev.\ D {\bf 47}, 3421
%(1993)
 [arXiv:hep-th/9211061];
%%CITATION = HEP-TH 9211061;%%
%}
%
%\TseytlinPQ
%\lref\TseytlinPQ{ A.~A.~Tseytlin,
``String vacuum backgrounds with covariantly constant null Killing
vector and 2-d quantum gravity,''
%Nucl.\ Phys.\ B {\bf 390}, 153 (1993)
 [arXiv:hep-th/9209023];
%%CITATION = HEP-TH 9209023;%%
``Exact string solutions and duality,'' arXiv:hep-th/9407099; }

%\MarolfBX
\lref\MarolfBX{ D.~Marolf and L.~A.~Zayas, ``On the singularity
structure and stability of plane waves,'' arXiv:hep-th/0210309.
%%CITATION = HEP-TH 0210309;%%
}

\lref\brecher{D. Brecher, J. P. Gregory, and P. M. Saffin,
``String Theory and the Classical Stability of Plane Waves, ''
arXiv:hep-th/0210308.}

\lref\brinkmann{H.W.Brinkmann, Math. Ann. 94 (1925) 119. }

%ENDREFERENCES

%%%%%%%%%%%%%%%%%%%%%%%
%kkkkkkkkkk

%%%%%%%%%%%%%%%%%%%%%%%
%1111111111

\Title{\vbox{\baselineskip12pt\hbox{hep-th/0212223}
\hbox{SU-ITP-02/47}}} {\vbox{
\centerline{Stringy Resolutions of Null Singularities}
%\bigskip
%\centerline{of}
%\bigskip
%\centerline{ Singularities}
 }}
\bigskip
\bigskip
\centerline{Michal Fabinger and Simeon Hellerman}
\bigskip
\centerline{ \sl Department of Physics and SLAC}

 \centerline{ \sl Stanford University}

 \centerline{ \sl Stanford, CA 94305/94309}

%\centerline{$^{1}${\it Department of Physics, Stanford University,
%Stanford, CA 94305}}
%\smallskip \centerline{$^{2}${\it SLAC Theory Group, MS 81, PO
% Box 4349, Stanford, CA 94309}}

\bigskip
\bigskip
\noindent

We study string theory in supersymmetric time-dependent
backgrounds. In the framework of general relativity, supersymmetry
for spacetimes without flux implies the existence of a covariantly
constant null vector, and a relatively simple form of the metric.
As a result, the local nature of any such spacetime can be easily
understood. We show that we can  view any such geometry as a
sequence of solutions to lower-dimensional Euclidean gravity. If
we choose the lower-dimensional solutions to degenerate at some
light-cone time, we obtain null singularities, which may be thought of as
generalizations of the parabolic orbifold singularity. We find that
in string theory, many such null singularities get repaired by
$\alpha'$-corrections - in particular, by worldsheet instantons.
As a consequence, the resulting string theory solutions do not
suffer from any instability. Even though the CFT description of
these solutions is not always valid, they can still be well
understood after taking the effects of light D-branes into
account; the breakdown of the worldsheet conformal
field theory is purely gauge-theoretic,
not involving strong gravitational effects.

\bigskip
\Date{December 2002}

%%%%%%%%%%%%%%%%%%%%%%%%%%%%%%%%%%%%%%%%%%%%%%%%%%%%%%%%%%%%%%%%%
%2222222222
\newsec{Introduction}

One of the most interesting features of string theory is its
ability to describe certain singular spacetimes, whose description
in general relativity inevitably breaks down. We have learned
much about the rich subject of static singularities and
their resolutions (for a review see {\it e.g.} \refs{ \Polchinski
\AspinwallMN \GreeneCY - \AspinwallFD }). Little is known,
however, about time-dependent singularities.

Recently, there has been an interesting attempt \LiuFT\ to
understand  the null singularity of the parabolic orbifold
 of Minkowski space \HorowitzAP, where the orbifold group is
generated by a parabolic element of the Lorentz group. Even though
the parabolic orbifold is a limit of a well-behaved string
background \refs{\LiuKB
 \FabingerKR - \HorowitzMW}, namely the null-brane
 \refs{\Figueroa,\ \SimonMA}, in the singular limit the backreaction of (almost) any
particle on the geometry becomes large, and perturbation theory
breaks down \refs{\LawrenceAJ, \LiuKB
 \FabingerKR - \HorowitzMW}.  
(Recently, there has been a lot of interest in time-dependent
backgrounds in general \refs{
 \KhouryBZ \BalasubramanianRY \CornalbaFI \NekrasovKF
\Figueroa \SimonMA  \LawrenceAJ \KoganNN \GutperleAI \BuchelWF
\aharony \ElitzurRT \CornalbaNV \CrapsII   \KachruKX \MartinecXQ
\CaiMR  \BuchelKJ \HemmingKD \HashimotoNR \SimonCF \AlishahihaBK
\SatohNJ \DolanPX \CaiSV \HubenyPJ \BachasQT \brecher \MarolfBX
\MyersBK \OkuyamaPC \AlishahihaBC \PapadopoulosBG \ChoGA -
\GauntlettFZ}. For other related work see {\it e.g.} \refs{
\AmatiSA \HorowitzBV \deVegaNR \TseytlinABC \NappiKV \KiritsisJK -
\KiritsisIJ }.)

The parabolic orbifold can be thought of as a circle fibration
over a nine-dimensional Minkowski space, with the fiber shrinking
along a null direction to a zero size. In this paper, we will be
interested in more general cases, with a smaller (but nonzero)
amount of supersymmetry, where more general fibers shrink
along a null direction in a slightly more general way. To
achieve this in a solution of general relativity or string theory,
we will have to consider also more general base spaces -- instead
of the Minkowski space, the fibrations will be over plane waves.

Some of the solutions constructed in this way will not have  a
much better behavior in string theory than the parabolic orbifold.
We will see, however, that decreasing the amount of supersymmetry
has rather dramatic consequences -- there are many distinct null
singularities of general relativity which are perfectly
well-behaved within the framework of string theory! In other
words, the ill-behaved singularities are exceptions, rather than
generic cases.

The paper is organized as follows. In section 2, we review some
basic facts about the parabolic orbifold singularity. In section
3, we discuss the properties of general purely geometric solutions
to supergravity which have a covariantly constant spinor (and
therefore also a covariantly constant vector). In section 4, we
consider a special case of these solutions, namely supersymmetric
fibrations over plane waves in general relativity. In section 5,
we provide a string theory description of these fibrations. In
section 6, we resolve null singularities. In conclusions, we
conclude. In  appendix A, we setup the coordinate system used in
section 3, and we prove some of its important properties. In
appendix B, we discuss the stability issues.

%%%%%%%%%%%%%%%%%%%%%%%%%%%%%%%%%%%%%%%%%%%%%%%%%%%%%%%%%%%%%%%%%%%%%%%%%%%%%
\newsec{A Warm-up Example: The Parabolic Orbifold Singularity}
%%%%%%%%%%%%%%%%%%%%%%%%%%%%%%%%%%%%%%%%%%%%%%%%%%%%%%%%%%%%%%%%%%%%%%%%%%%%%
% aaaaaaaaaa

Before get to the case of more general null singularities, let us
briefly review the properties of one of the most simple null
singularities -- the parabolic orbifold singularity. Although we
do not know how to resolve it in string theory, it will provide a
useful intuition for a class of null singularities which do have a
non-singular behavior in string  theory.

%%%%%%%%%%%%%%%%%%%%%%%%%%%%%%%%%%%%%%%%%%%%%%%%%%%%%%%%%%%%%%%%%%%%%%%%%%%%%%%
\subsec{The Parabolic Orbifold of Minkowski Space}

This orbifold can be obtained from a three-dimensional Minkowski
space $\IR^{1,2}$ (cross $\IR^{7}$ in case we want to consider
superstring theory) by modding out by a group isomorphic to \IZ\
and generated by a parabolic element of $SO(1,2)$:
\eqn\generator{{\sl g}_0 = e^{i \beta J}, \quad J \equiv {1 \over
\sqrt{2}}\ J^{01} + {1\over \sqrt{2}}\ J^{12}.}
In terms of the coordinates
\eqn\coord{\rmx^+ = {\rmx^0 + \rmx^1\over \sqrt{2}},\quad  \rmx^-
= {\rmx^0 - \rmx^1\over \sqrt{2}}, \quad \rmx = \rmx^2,}
the generator \generator\ acts as
\eqn\orbifoldaction{ \pmatrix{ \rmx^+ \cr  \rmx^-\cr \rmx^{~}\cr}
\quad\rightarrow\quad \pmatrix{ \rmx^+  \cr \rmx^- + \beta \rmx +
\half\beta^2 \rmx^+ \cr \rmx + \beta \rmx^+ } }
%%
%
%\eqn\orbifoldaction{ \pmatrix{ x^+ \cr x^{~}\cr x^-\cr \chi \cr}
%\quad \rightarrow\quad \pmatrix{ x^+ \cr x + v x^+  \cr x^- + v x
%+ \half v^2 x^+ \cr \chi  + L\cr} }
 If we introduce a new set of
coordinates
\eqn\ycoordinates{\eqalign{ u & = \rmx^+ \cr
  v &  = \rmx^- -  {\rmx^2\over 2\rmx^+} \cr
   x    & =  {\rmx\over \rmx^+},   \cr
}}
the orbifold identifications \generator\ become very simple:
 \eqn\yidentifications{(u,  v , x ) \sim (u, v, x +\beta  ),}
and the metric can be written in a the following form
 \eqn\ymetric{ ds^2  = -2\ du dv + u^2 dx^2. }
Strictly speaking, the definition of $v$ and $x$ in \ycoordinates\
is sensible only for non-zero $u$. The slice of $u = 0$
corresponds to a null singularity, close to which the full
orbifold spacetime is not even Hausdorff.

If we interpret the coordinate $u$ as the light-cone time, we can
view the region of negative $u$ as a light-cone-time evolution of
a shrinking circle. Its circumference is given by $\beta u$, and
in particular, for $u = 0$ the circle degenerates to a zero
size.\foot{It is also possible to consider more general null
orbifolds which can be interpreted as $d$-dimensional tori
shrinking along $u$. These orbifolds can be obtained by modding
out the Minkowski space by $\IZ^d$ generated by exponentials of
$J^{0i} + J^{1i}$ for $i=1..d$, in the rectangular case.}

This  singularity does not seem to be better-behaved in string
theory than in general relativity. It has been argued that adding
just a single particle (with non-zero $p^u$) into the orbifold
causes so large backreaction that the approximation of small
perturbations around the background geometry fails, and in
particular, the string perturbation theory is invalid. One can see
this effect also directly from the singular behavior of the string
scattering amplitudes.

One important feature of the geometry \ymetric\ is that in a
certain sense, it represents an infinite distance in the moduli
space of circles, traversed in a finite
light-cone time. (This follows from the fact that the zero size
circle is infinitely far from any other point in the moduli space
of the $S ^1$, whether or not we include the $\ap$ corrections
of string theory).
 For this reason it will be interesting to consider more
general null singularities, where the circle is replaced by some
different internal space whose zero volume limit is not an
infinite distance from the rest of the moduli space.

%%%%%%%%%%%%%%%%%%%%%%%%%%%%%%%%%%%%%%%%%%%%%%%%%%%%%%%%%%%%%%%%
\newsec{General Properties Of Null Geometries With A Constant Spinor}
%%%%%%%%%%%%%%%%%%%%%%%%%%%%%%%%%%%%%%%%%%%%%%%%%%%%%%%%%%%%%%%%
%bbbbbbbbbb

In this section, we will be interested in the properties of
general solutions to ten-dimensional supergravity which have at
least one conserved Majorana-Weyl spinor, assuming that there are
no fluxes and that the dilaton is constant. We will choose the
background metric in the 10d `string' frame to be the same as in
the 10d Einstein frame. This is possible because we will keep the
dilaton fixed. (The condition of a constant dilaton can be relaxed
quite easily. We will not do so in order to keep the discussion
relatively simple.)

The geometries of this type will necessarily admit a covariantly
constant null vector\foot{
This means that they will belong to the family of plane-fronted
waves with parallel rays (pp-waves), because pp-waves are {\it
defined} to be spacetimes with a covariantly constant null vector.
Note also that if a vector is covariant constant, it is also a
Killing vector.
}, which can be seen as follows. With the simplifying assumptions
above, unbroken supersymmetry implies \refs{\Polchinski, \GreenSP}
that there exists some number of covariantly constant
Majorana-Weyl spinors $\eta^{(I)}$, \it i.e. \rm spinors
satisfying \ee{ \G \eta^{(I)} = + \eta^{(I)}\hskip .3in
\eta^{(I)}{}^* = C^* \eta^{(I)}, } where $\G$ and $C$ are the 10D
chirality and charge conjugation matrices, respectively, and the
star denotes complex conjugation.  Now, we can construct
covariantly constant vector fields as linear combinations of
$\bar\eta^{(I)}\Gamma^\mu \eta^{(J)}$, which will be symmetric in
$I$ and $J$, since the matrices $(C^*\G^0\G^\m)_{\a\b}$ are
symmetric in the spinor indices $\a, \b$.  The number of
independent vector fields obtained in this way will be
non-zero\foot{
This is a consequence of working in a spacetime of Lorentzian
signature. For instance in the case of six-dimensional Euclidean
supergravity on a generic Calabi-Yau three-fold, having a
covariantly constant spinor does not imply the existence of any
Killing vectors.
}, since for example for the $I$-th spinor,
$\eta^{\dagger(I)}\Gamma^0 \Gamma^\mu \eta^{(I)} = 0$ would imply
$\eta^{\dagger(I)} \Gamma^0 \Gamma^0 \eta^{(I)}=\eta^{\dagger(I)}
\eta^{(I)}  = 0$, and consequently $\eta^{(I)} = 0$. Furthermore
we can show that every vector $l^{\mu(I)}\equiv
\bar{\eta}^{(I)}\Gamma^0 \Gamma^\mu \eta^{(I)}$ is light-like: the
quantity \ee{ (C^*\G^0\G^\m)_{\a\b}(C^*\G^0 \G_\m)_{\g\d} + {\rm
permutations ~of}~ \a,\b,\g,\d } vanishes\foot{See for example pg.
246 of \GreenSP}, and contracting this  into $\eta^{(I)}_\a
\eta^{(I)}_\b \eta^{(I)}_\g \eta^{(I)}_\d$ shows that $\l^{\m(I)}
\l_{\m}^{(I)} = 0.$ We will be, of course, most interested in the
case when none of the linear combinations of $\l^{\m(I)}$ is
time-like. As a result, the covariantly constant null vector
$l^\mu$ will be unique up to a constant rescaling.

There are several important properties of these spacetimes, which
we prove (and refine) in  appendix A. First of all, we can define
a certain special coordinate $u$ with the property that the slices
of constant $u$ are light-like surfaces. Locally in $u$, {\it
i.e.} for some range $(u_{\rm a}, u_{\rm b})$, the geometry can be
written as a fibration over a null geodesic with affine parameter
$u$.
Now, if we start at any non-singular fiber, then locally in $u$,
we can always find coordinates $v$ and $x^i$, $i=1..8$, such that
the metric becomes\foot{
If the fibers are topologically non-trivial, we might need to use
more  patches of coordinates $x^{i}_{(a)}$, labelled by an index
$a$. In order to keep the notation simple, we will not always
explicitly mention this fact. The coordinate $v$ will be, however,
always globally well-defined. We will assume that $v$ is
non-compact, because for a compact $v$, there  would be closed
causal curves through every point.
}
\eqn\ppmetric{ds_{10}^2 = -2\ \! du\ \! dv + h_{ij}(u,x^k)\  \!
dx^i\ \! dx^j.}
Here, $x^i$ parameterize a space which can be either compact or
non-compact. In these coordinates the constant vector $l^\mu$ can
be written as  $l^u =0,\ l^v = 1, \ l^{i} = 0$. The metric
\ppmetric\ does not depend on $v$, which is a consequence of
$l^\mu$ being a Killing vector.

Note that the metric of the parabolic orbifold \ymetric\ is
precisely of the form \ppmetric, with $x^i$ parameterizing $S^1
\times \IR^7$ and $h_{11}=u^2$, $h_{ij}= \d_{ij}$ for $i,j=2..8$.
(In \ymetric, the directions 2..8 were suppressed.)

%%%%%%%%%%%%%%%%%%%%%%%%%%%%%%%%%%%%%%%%%%%%%%%%%%%%%%%%%%%%%%%%%
\subsec{Slide-Show Interpretation of the Spacetimes}

 The metric \ppmetric\ has very simple transformation
properties under boosts. If we perform a coordinate change
\eqn\boost{ u = u_0 + \Omega( u' - u_0), \quad v = \Omega^{-1}
 v', }
it becomes
\eqn\slowmetric{ds_{10}^2 = -2\ \! du'\ \! dv' + h_{ij}(u_0 +
\Omega u' - \Omega u_0,x^k)\ \! dx^i\ \! dx^j.}
This is just a manifestation of the Doppler effect for
gravitational waves. We see that by choosing $\Omega$ to be small,
we can make the metric arbitrarily slowly varying. In other words,
where is no scale associated with the $u$-dependence of the
metric. In the strict limit $\Omega \to 0$, we obtain
\eqn\frozenmetric{ds_{10}^2 = -2\ \! du'\ \! dv' +
h_{ij}(u_0,x^k)\ \! dx^i\ \! dx^j.}
If the original metric satisfied Einstein's equations, then the
frozen metric \frozenmetric\ must also satisfy them. The spacetime
\frozenmetric\ has the form of a direct product of a
two-dimensional Minkowski space, and an eight-dimensional space
parameterized by $x^i$. This implies that $h_{ij}(u_0,x^k)$  must
be a solution of 8d Euclidean gravity for any fixed $u_0$. (It is
clear that if the original solution \ppmetric\ was supersymmetric,
 then also these 8d solutions must admit a covariantly
constant spinor.)

As a result, the spacetime \ppmetric\ can be interpreted as a
series of slices of constant $u$, where each slice is a direct
product of a line, parameterized by $v$, and a solution to
eight-dimensional Euclidean gravity. The precise metric on these
slices will vary with the light-cone time $u$.

It is natural to ask whether any path in the space of
(supersymmetric) solutions of 8d Euclidean gravity gives rise to a
solution to 10d Einstein's equations in this way. We will see in
the following sections (in a slightly less general context) that
this is not the case, and that there is one additional condition
on such path that has to be satisfied.

%%%%%%%%%%%%%%%%%%%%%%%%%%%%%%%%%%%%%%%%%%%%%%%%%%%%%%%%%%%%%%%%
\newsec{Supersymmetric Fibrations Over Plane Waves}
%%%%%%%%%%%%%%%%%%%%%%%%%%%%%%%%%%%%%%%%%%%%%%%%%%%%%%%%%%%%%%%%
%cccccccccc

The null singularities we will be eventually most interested in
correspond to some compact $\tilde d$-dimensional spaces $\CM$
(with metric $ h_{ab}$) collapsing to a zero size at some
light-cone time $u_s$. For simplicity, we will concentrate on the
case where \ppmetric\ takes the form
\eqn\cywave{ds_{10}^2 = -2\ \! du\ \! dv + a^2(u) \ \! dy^{\alpha}
\ \! dy^{\alpha} + h_{ab}(u,x^c)\ \! dx^a\ \! dx^b . }
In other words, we will split the $x^i$ coordinates into
$y^\alpha$, $\alpha = 1..(d-2)$, which will be coordinates on a
flat $(d-2)$-dimensional plane with a $u$-dependent scale factor,
and $x^a$, $a=1...\tilde d$, parameterizing the compact space
$\CM$. Here $d=10-\tilde d$. The section of the spacetime spanned
by $u, v$, and $y^\alpha$ has the most simple form possible in
this context -- it is a $d$-dimensional plane wave.
By the argument from the previous section, we see that at any
fixed $u_0$, the compact manifold $\CM$ has to be of special
holonomy.

\subsec{Kaluza-Klein reduction}

Now we can perform a Kaluza-Klein decomposition of the metric
\cywave\ in order to obtain a lower-dimensional description. The
spacetime \cywave\ can be viewed as an $\CM$-fibration over a
plane wave. Since the fiber $\CM$ is everywhere orthogonal to the
base, the KK gauge fields will be zero, and the lower-dimensional
metric in the `string' frame will be simply equal to the metric on
the base,
\eqn\planewave{ds_{d,s}^2 = -2\ \! du\ \! dv + a^2(u) \ \!
dy^{\alpha} \ \! dy^{\alpha}. }
This is a plane wave metric in the usual Rosen coordinates.
There will be also some number
of $d$-dimensional effective scalars $\phi^A(u)$, $A=1...n$,
corresponding to the moduli of $\CM$. In particular, there will be
a certain function $\omega$ of the scalars describing the total
volume of $\CM$, $V= V_0\ {\rm exp }{(\omega)}$.

If we want to express the metric \planewave\ in the
$d$-dimensional Einstein frame, we have to perform a Weyl
rescaling.
\eqn\einsteinmetric{ds_{d,{\rm E}}^{2}= \left( {V(u)\over V_0}
\right)^{2 / (d-2)} \cdot \ ds_{d,{\rm s}}^{2}= {\rm exp}\left(
{2\omega(u)\over d-2} \right) \ ds_{d,{\rm s}}^{2}.}
When $V(u)<V_0$ the distances look shorter in the Einstein frame
than in the `string' frame. By a simple reparameterization of $u$
and a redefinition of $a$, we can put the metric \einsteinmetric\
into the Rosen form even in the Einstein frame
\eqn\planewaveeinstein{ds_{d,{\rm E}}^2 = -2\ \! d\tilde u\ \! dv
+ \tilde a^2(\tilde u) \ \! dy^{\alpha} \ \! dy^{\alpha}. }
More explicitly
\eqn\redefinition{\tilde u = \int {\left({V(u)\over
V_0}\right)}^{2 / (d-2)} du, \quad \tilde a(\tilde u) =
{\left({V(u)\over V_0}\right)}^{1 / (d-2) } a(u).}

\subsec{Equations of motion}

The equations of motion for the ten-dimensional {\it background}
\cywave\ are equivalent to the $d$-dimensional Einstein's
equations for the metric \einsteinmetric, \planewaveeinstein
\eqn\einstein{R_{\mu \nu}^{( E)} - \half R^{( E)} g_{\mu \nu}^{(
E)} = 8\pi T_{\mu \nu}^{( E)}.}
The energy-momentum tensor is sourced by the minimally coupled
scalars $\phi^A$, and can be expressed as
\eqn\stresstensor{T_{\mu \nu}^{( E)} = {\cal G}_{AB}\ \! \p_\mu
\phi^A \p_\nu \phi^B - \half g_{\mu \nu}^{( E)}\ \! {\cal G}_{AB}\
\! \p_\sigma \phi^A \p^\sigma \phi^B ,}
where ${\cal G}_{AB}(\phi^C)$ is the metric on the moduli space of
${\CM}$. In our case, the scalars $\phi^C$ depend only on $\tilde
u$. As a result, the second term in \stresstensor\ vanishes, and
the only non-zero component of the energy-momentum tensor is
\eqn\stresstensortwo{T_{\tilde u \tilde u}^{( E)} = {\cal
G}_{AB}(\phi^C)\ \! \dot \phi^A \ \! \dot \phi^B,}
where the dots denote differentiation with respect to $\tilde u$.
The Einstein tensor on the left-hand side of \einstein\ also takes
a simple form, the only non-zero component being
\eqn\einsteintwo{G_{\tilde u \tilde u}^{( E)} =  R_{\tilde
u \tilde u}^{( E)} =
{(d-2)}\ {\ddot {\tilde a} \over {\tilde a}}\ . }
We see that in our case, Einstein's equations for the metric
\cywave\ reduce just to a single equation,
\eqn\constraint{ 8\pi {\cal G}_{AB}(\phi^C)\ \! \dot \phi^A \ \!
\dot \phi^B = {(d-2)}\ {\ddot {\tilde a} \over {\tilde a}}\ . }
%
%%
%\eqn\constraint{ 8\pi {\cal G}_{AB}(\phi^C)\ \! \dot \phi^A \ \! \dot
%\phi^B = - \ddot \omega + {\dot \omega^2\over d-2 } + {(d-2)}\
%{\ddot a \over a}. }
%
Given a path $\phi^A(\tilde u)$ in moduli space, one can always
solve locally in $\tilde u$ for the scale factor $a(\tilde u)$.

The equations of motion for the scalars are trivial: Assuming that
the scalars depend just on $\tilde u$, they will be automatically
satisfied.

\subsec{Brinkmann coordinates for the plane wave}

For completeness, let us  mention that there is also another
useful set of coordinates for the plane-wave part of \cywave,
known as the Brinkmann coordinates \brinkmann. By a
redefinition of $v$ and $y^\alpha$, which is well known in the
theory of plane waves \ppwaves, we can put
the metric \cywave\ into the following form
\eqn\cywavebrinkman{ds_{10}^2 = -2\ \! du\ \! d{\rm v} + b(u)\ \!
{\rm y}^{\alpha} {\rm y}^{\alpha} \ \! du^2  + \ \! d{\rm
y}^{\alpha} \ \! d{\rm y}^{\alpha} + h_{ab}(u,x^c)\ \! dx^a\ \!
dx^b . }
Similarly, the $d$-dimensional metric \planewaveeinstein\  can be
written as
\eqn\planebrinkmaneinstein{ds_{d,{\rm E}}^2 = -2\ \! d\tilde u\ \!
d{\rm v} + \tilde b(\tilde u)\ \! \tilde {\rm y}^{\alpha} \tilde
{\rm y}^{\alpha} \ \! d\tilde u^2 + \ \! d\tilde {\rm y}^{\alpha}
\ \! d\tilde {\rm y}^{\alpha}. }
The equation of motion \constraint\ then becomes
\eqn\constraintbrinkman{8\pi {\cal G}_{AB}(\phi^C)\ \! \dot \phi^A
\ \! \dot \phi^B = {(d-2)}\ \tilde b. }
The advantage of the Brinkmann coordinates is that they can cover
all of the plane wave without ever degenerating.

\subsec{Supersymmetry}

Any path in the moduli space of  $\CM$ leads to a supersymmetric
gravitational solution of the type \cywave, \cywavebrinkman, as
long as we satisfy Einstein's equations \constraint,
\constraintbrinkman\ by an appropriate choice of $\tilde a$ or
$\tilde b$. The fact that the solution will be supersymmetric can
be seen from the $d$-dimensional description. As we said, in
$d$-dimensions, the metric will be of the plane wave form, and the
only other non-zero fields will be the effective scalars $\phi^A$.
It is known that  every plane wave admits a covariantly constant
spinor. More explicitly, its existence can be seen as follows.

Let us work in the globally well-behaved Brinkmann coordinates
\planebrinkmaneinstein. If we choose the vielbein to be
$e^{(\tilde u)} = d\tilde u, \ e^{({\rm v})} = d{\rm v}-\half
\tilde b(\tilde u)\ \! \tilde {\rm y}^{\alpha} \tilde {\rm
y}^{\alpha}\ \! d\tilde u, \ e^{(\alpha)} = d\tilde {\rm
y}^\alpha$, we can express the covariant derivatives for the
spinors as $\nabla_{\tilde u} = \partial_{\tilde u} - {1 \over 2}
\tilde b(\tilde u) \ \! \tilde {\rm y}^\alpha \ \! \Gamma_{\rm v}
\Gamma_\alpha, \nabla_{\rm v} =
\partial_{\rm v},  \nabla_{\alpha} =
\partial_{\alpha}$. Clearly, any spinor $\epsilon$ which is constant for
this choice of vielbein and which satisfies $\Gamma_{\rm
v}\epsilon = 0$ will be also covariantly constant,
$\nabla_\mu\epsilon=0 $.

Now it is not hard to see that the supersymmetry generated by any
such spinor is preserved by the background we consider. The
supersymmetric variation of the gravitino obviously vanishes
because it is proportional to the covariant derivative of the
spinor, $\nabla_\mu\epsilon$. Similarly, the variations of
spin-$\half$ fermions vanish, because the only possible
non-trivial terms all contain $\Gamma^\mu \partial_\mu \phi^A
(\tilde u) \ \! \epsilon$, which is zero due to $\Gamma_{\rm v
}\epsilon = 0$.

\subsec{Conclusion}

Any supersymmetric spacetime of the form \cywave, or
\cywavebrinkman, can be thought of as a path in the moduli space
of a special-holonomy manifold $\CM$, which is parameterized by
$u$ (resp. $\tilde u$). Conversely, if we choose any such path, we
can always construct a gravitational solution of the form \cywave,
or \cywavebrinkman, because the only non-trivial Einstein equation
\constraint, \constraintbrinkman\ can be easily solved by a
suitable choice of $\tilde a(\tilde u)$, or $\tilde b(\tilde u)$.
Any solution constructed in this way will be supersymmetric.

%%%%%%%%%%%%%%%%%%%%%%%%%%%%%%%%%%%%%%%%%%%%%%%%%%%%%%%%%%%%%%%%%
\newsec{Stringy  Description Of Supersymmetric Fibrations Over Plane Waves}
%%%%%%%%%%%%%%%%%%%%%%%%%%%%%%%%%%%%%%%%%%%%%%%%%%%%%%%%%%%%%%%%%

%dddddddddd

To obtain a string theory description of the fibrations over plane
waves studied in the previous section, one  might try to write
down a non-linear sigma model based on the target space metric
\cywave, \cywavebrinkman, and then  correct it order by order in
$\ap$ to obtain a conformal field theory. However, since we are
interested in  cases in which the size of the fiber becomes of
order the string length and where the non-linear sigma model
perturbation theory breaks down, we would not get too far in this
way.

Let us first discuss string theory counterparts of non-singular
solutions to general relativity \cywave, \cywavebrinkman,
postponing the discussion of the singular cases until section 7.
To get a handle on the string theory description of the spacetimes
\cywave, \cywavebrinkman, we will use their transformation
properties under boosts. As discussed in section 3, by performing
a large enough boost, we can make the $u$-dependence (resp.
$\tilde u$-dependence) of the solutions arbitrarily slow.
Alternatively, we can simply {\it start} with a slowly varying
spacetime. For such slowly varying geometry, the low-energy
effective description of the spacetime will be perfectly valid.

%More precisely, the $d$-dimensional effective field theory description
%of the system could break down in various ways:
%
%$\bullet{}$ The effective scalars in $d$-dimensions could vary quickly,
%or the $d$-dimensional curvatures could become large.
%
%$\bullet{}$ The nonzero modes on the internal space could become light.
%
%$\bullet{}$ There could be other degrees of freedom, such as
%winding strings or wrapping branes, which become light.%
%
%
%The second and third possibilities can be avoided, by choosing the path
%$\phi^A(u)$ in moduli space to avoid point in moduli space where
%the light spectrum is enhanced.   Or if we choose not to avoid these
%points, we merely encounter breakdowns of lower dimensional field theory
%which are quite well understood from the case of static backgrounds.
%The first of the three possibilities
%can be, in
%essence, boosted away, as long as the $u$-derivatives of the scalars
%%$\phi^{\prime A}(u)$ along moduli space are bounded.  It is therefore
%easy to set up conditions such that the d-dimensional effective field
%theory describes the entire solution reliably.  We will stress that
%this may happen even if the classical supergravity solution is singular, as
%$\ap$-corrections will often change infinite gradients on moduli
%space into finite ones!

This reduces the problem of finding the
 string theory counterpart of \cywave,
 \cywavebrinkman\
 to the well-studied problem of
finding  moduli space metrics for ordinary string theory
compactifications. The lower-dimensional metric will be given by
the same expressions as before \planewaveeinstein,
\planebrinkmaneinstein.
%%
%\eqn\stringeinsteinmetric{ds_{d,{\rm E}}^{2}= e^{{2\omega(u)/
%(d-2)} } \ ds_{d,{\rm s}}^{2},  }
%%
%where \eqn\stringplanemetric{ds_{d,s}^2 = -2\ \! du\ \! dv +
%a^2(u) \ \! dy^{\alpha} \ \! dy^{\alpha}, }
%%
%or
%%
%\eqn\stringplanemetrictwo{ds_{d,s}^2 = -2\ \! du\ \! d{\rm v} +
%b(u)\ \! {\rm y}^{\alpha} {\rm y}^{\alpha} \ \! du^2 + \ \! d{\rm
%y}^{\alpha} \ \! d{\rm y}^{\alpha}.}
%%
Also the stringy equations of motion will look very similar to
\constraint, \constraintbrinkman, namely
\eqn\stringconstraint{ 8\pi {\cal G}_{AB}^{(str)}(\phi^C)\ \! \dot
\phi^A \ \! \dot \phi^B = {(d-2)}\ {\ddot {\tilde a} \over {\tilde
a}}, }
 or
\eqn\stringconstrainttwo{8\pi {\cal G}_{AB}^{(str)}(\phi^C)\ \!
\dot \phi^A \ \! \dot \phi^B =  {(d-2)}\ \tilde b,}
with the difference that now we have to use the $\ap$-corrected
moduli space metric for string theory compactified on $\CM$. (If
we wish to work at non-zero string coupling, we should also
include the $g_s$-corrections.)

Since we have  boosted the original spacetime to make it very
slowly varying, it is clear that if there are any
$\ap$-corrections to the equations of motion besides those already
included in the moduli space metric ${\cal G}_{AB}^{(str)}$, they
must be rather small. It is interesting to note that actually any
such corrections which are perturbative in $\ap$  vanish
identically for our solution. This is because perturbative
corrections to the equations of motion correspond to some
higher-derivative terms added to the Einstein equations \einstein.
They have to be generally covariant, and in our case, they have to
be made from the lower-dimensional metric and the scalars. Since
the metric \planewaveeinstein\ and the scalars $\phi^A$ depend
only on $\tilde u$, constructing a non-vanishing
 tensor with two free indices and more than two
derivatives requires contracting at least two $\tilde u$-indices.
Such a contraction, however, makes any tensor vanish, because the
corresponding metric coefficient is zero. We see that the
symmetries of the problem, and in particular the absence of any
scale associated with the $\tilde u$-dependence of the solution,
forbid any further perturbative $\ap$-corrections to
\stringconstraint.   This argument generalizes that of
\HorowitzBV\ to the case where the low-energy effective dynamics
of string theory are those of gravity coupled to scalar fields.
% The absence of such scale also suggests that
%this is true even for any additional non-perturbative
%$\ap$-corrections.

\subsec{CFT description}

Even in static cases it is hard to get an explicit lagrangian for
the worldsheet CFT when the  curvature of the target space becomes
of order the string scale. To analyze such compactifications, one
has to rely on some less direct methods.
%
% such as the linear sigma
%model for the worldsheet CFT \WittenYC.
%One constructs a two-dimensional linear sigma model which in some
%range of parameters flows in the infrared into the non-linear
%sigma model with the target space of interest. Varying parameters
%of the linear sigma model allows one to change the moduli of the
%target space. At some point, when the target space curvature
%reaches the string scale, the non-linear sigma model perturbation
%theory breaks down.  Even when this happens, the linear sigma
%model generically defines a well behaved CFT as its infrared fixed
%point \WittenYC. By studying the linear sigma model one can learn
%a great deal about the properties of string theory in the regime
%where the internal space has small volume.
%
It is clear that in general, we will not be able to write down
explicitly the CFT action providing a string theory description of
the spacetimes \cywave, \cywavebrinkman\ when the fiber becomes
small. We can, however, at least write down its general form. Let
us denote $\CL[\psi^K; \phi^A]$ the worldsheet lagrangian which
corresponds to the space $\CM$ at some fixed values $\phi^A$ of
its moduli and  which functionally depends on some worldsheet
fields $\psi^K$. Now, the worldsheet action describing the
spacetimes of interest can be written schematically as
\eqn\cftaction{S = -{1 \over 4\pi \ap} \int d^2\sigma \ \!
\sqrt{-\gamma} \ \! \left( -2 \ \! \p_a u\ \! \p^a {\rm v} + b(u)\
\! {\rm y}^{\alpha} {\rm y}^{\alpha} \ \! \p_a u\ \! \p^a u + \ \!
\p_a {\rm y}^\alpha \ \! \p^a {\rm y}^\alpha \ \! + \CL[\psi^K;
\phi^A(u)] \right),}
where we have ignored all the fermions not contained in $\CL$. We
should stress that here, $\phi^A$ are not independent worldsheet
fields, but merely some functionals of $u$ related to $b(u)$ by
\stringconstrainttwo. If we were powerful enough, this CFT would
allow us, in principle, to compute  string scattering amplitudes
beyond the low-energy field theory approximation.

%%%%%%%%%%%%%%%%%%%%%%%%%%%%%%%%%%%%%%%%%%%%%%%%%%%%%%%%%%%%%
\newsec{Generalizations To Geometries With Non-Trivial Potentials And Fluxes}
%%%%%%%%%%%%%%%%%%%%%%%%%%%%%%%%%%%%%%%%%%%%%%%%%%%%%%%%%%%%%
%eeeeeeeeee

Most of what we said about fibrations over plane waves has a
straightforward generalizations  to the cases where the fiber
$\CM$ carries some non-trivial potentials and fluxes (with the
exception of the worldsheet point of view, since Ramond-Ramond
fields or non-trivial dilaton are usually problematic for string
perturbation theory in general).  All we have to do is to replace
the moduli space of metric ${\cal G}_{AB}$ on $\CM$ by an
appropriate moduli space of the desired string theory
compactifications. It is not clear to us, however, whether there
is also any simple generalization of the statement that in
supergravity, any purely geometric solution with null
supersymmetry takes locally the form \ppmetric.

%%%%%%%%%%%%%%%%%%%%%%%%%%%%%%%%%%%%%%%%%%%%%%%%%%%%%%%%%%%%%
\newsec{Stringy Resolutions Of Null Singularities}
%%%%%%%%%%%%%%%%%%%%%%%%%%%%%%%%%%%%%%%%%%%%%%%%%%%%%%%%%%%%%
%ffffffffff

We have seen that in general relativity the local structure of
spacetimes of the form \cywave, \cywavebrinkman\ can be understood
in terms of paths in the moduli space of the compact manifold
$\CM$. If we choose the path to reach (in finite $\tilde u$) the
boundary of the classical moduli space where the $\CM$ shrinks to
a zero size, the spacetime will have a null singularity which can
be thought of as a generalization of the parabolic orbifold
singularity. At this point, general relativity certainly breaks
down. Moreover, it seems that (almost) any particle added to the
spacetime would make the singularity spacelike, essentially
because  a finite amount energy would be focused into an
infinitely small region.

It seems that string theory is too weak to change anything
substantial  in this kind of story. In the case of the parabolic
orbifold of Minkowski space, we have seen very well how string
theory loses its fight against the null singularity! Unless we
choose a slightly different (non-singular) classical solution to
begin with, we do not know, at the present time, how to deal with
such spacetime.

This is all true, but the reason why this happened is that we were
 really harsh. We constrained the string theory by such a
large amount of supersymmetry that it could not protect itself by
using one of its most powerful weapons -- the worldsheet
instantons! If we decide to reduce the amount of supersymmetry, it
will have extremely dramatic consequences.

\subsec{General considerations}

We will consider null singularities which arise when the fiber
$\CM$ shrinks to a zero size in finite $\tilde u$. We do not have to
assume that the ten-dimensional dilaton is necessarily constant,
and also, we may allow the space $\CM$ to carry non-trivial
potentials or fluxes.

In general relativity, the   moduli space distance to a
configuration of a vanishing volume is always infinite. String
theory offers more possibilities (assuming that we have a
reasonable definition of the volume even for $\CM$ of order the
string length):

In some cases, usually with a large amount of supersymmetry, there
are no important corrections to the moduli space metric and the
distance to the zero volume configuration remains infinite. If we
want to cover this infinite distance in a finite light-cone time
$\tilde u$, there will be no justification for the low-energy
approximation we have been using. Moreover, the zero volume limit
does not lie inside of the moduli space, but rather on its
boundary. It is not clear whether it is makes any sense to ask
what should happen once we reach this boundary. We do not know, at
present, how to study such singularities in a controllable way.

%It seems quite unlikely, at present, that this kind of
%singularities can be studied in any controllable way and that
%there are any effects which would make the stringy physics
%non-singular.  Amplitudes will almost certainly appear singular in
%string perturbation theory, as demonstrated beautifully and in
%great detail in \LiuKB, for the case of the parabolic orbifold.

On the other hand in more generic cases, the zero-volume limit is
either just a finite distance in the quantum-corrected moduli
space(\CandelasRM, \CandelasQD), or it does not exist at all. For
the corresponding string theory solutions, the lower-dimensional
description we have been using so far is perfectly valid. (At
certain points of the moduli space we might be forced to include
more fields into the lower-dimensional description.) For this
reason, there will be no instability similar to that of the
parabolic orbifold which was studied in detail in \HorowitzMW.
After all, the system we are considering is just a string theory
compactification moving arbitrarily slowly in its moduli space.

Rather than continuing this general discussion, let us now focus
on a more specific context.

\subsec{Calabi-Yau three-fold fibrations over plane waves}

If we compactify type IIA string theory on a Calabi-Yau
three-fold, we obtain in four dimensions an $\CN=2$ effective
field theory which contains one gravity multiplet (with no
scalars), $h^{1,1}$ vector multiplets (each containing two real
scalars, for example the overall volume modulus), and $h^{2,1}+1$
hypermultiplets (each containing four real scalars, for example
the dilaton). The moduli space of the whole theory exactly
factorizes into the vector multiplet moduli space and the
hypermultiplets moduli space (up to discrete quotients), and for
this reason we can consider these two spaces separately.

In particular, we will consider motions only in the vector
multiplet moduli space, since it is the vector multiplets which
control the K\" ahler parameters (including the overall volume) of
the Calabi-Yau manifolds. The vector multiplet moduli space metric
in type IIA receives no $g_s$-corrections, and in principle, it
could be determined from the classical contribution and the
contributions of the worldsheet instantons at zero string coupling
(there are no perturbative $\ap$-corrections). In practice, it
much more convenient to use the mirror map between $\CM$ and a
different Calabi-Yau manifold $\CW$ in type IIB, because in type
IIB,  the vector multiplet moduli space metric does not receive
any $\ap$ or $g_s$-corrections at all.

The overall quantum volume of the Calabi-Yau $\CM$ may be defined
to be equal to the mass of the D6-brane wrapping $\CM$
\refs{\GreeneTX, \PolchinskiSM}. Of course, for a large
Calabi-Yau, this definition coincides with the classical
definition of the volume.
%
%$\bullet$ \ There may be no point in the quantum-corrected moduli
%space of the Calabi-Yau compactification where the D6-branes
%wrapping the whole Calabi-Yau would  become massless.
%In this case, the singular solution of general
%relativity automatically gets
% replaced by a totally non-singular string theory
%solution. -- This is unlikely.
%
%
For a typical Calabi-Yau (or maybe in all cases), there is a
finite-distance point in the quantum-corrected moduli space  where
the D6-branes become massless. This means that we can construct
fibrations over plane waves where the fiber $\CM$ literally
shrinks to a zero size at some $u=u_s$. Because the zero-volume
point is a finite distance in the moduli space, we will not lose
any control over the solution. The lower dimensional effective
description will still be perfectly valid, it will just contain
new light degrees of freedom coming from the D6-branes.

%$\bullet$ \ In principle, there is also the possibility that the
%D6-branes become massless at an infinitely-distant point in the
%quantum-corrected moduli space, even though we do not know any
%examples of this kind (among Calabi-Yau three-folds). This kind of
%behavior can be, however, encountered for example in the case of
%tori. It is not clear how one would resolve the corresponding null
%singularities.
%This is also unlikely.

 \subsec{An Example: A
Shrinking Quintic }

The quintic hypersurface in $\CP^4$ (denoted $P_4(5)$) is given in
projective coordinates by the equation
\eqn\quintic{z_1^5 + z_2^5 + z_3^5 + z_4^5 + z_5^5 = 0}
(or by one of its possible deformations by other fifth order
monomials). We can  choose the metric on $P_4(5)$ to be
Ricci-flat, and $P_4(5)$ becomes a Calabi-Yau manifold. The Hodge
numbers which give rise to its moduli are $h^{1,1}=1$ and
$h^{2,1}=101$. This means that in type IIA, the moduli space of
vector multiplets (which control the K\" ahler parameters of the
quintic) will have complex dimension one. In other words, there
will be just one vector multiplet.

\ifig\quintic{The shaded region in this figure represents
schematically the quantum vector multiplet moduli space of the
quintic. The semi-infinite line going upwards from ${\cal P}_0$
should be identified with a similar line originating from ${\cal
P'}_0$. Similarily, ${\cal P}_0$-${\cal P}_{LG}$ is to be
identified with ${\cal P'}_0$-${\cal P}_{LG}$. There are three
important points in the moduli space: The zero-volume point ${\cal
P}_0$, the Landau-Ginzburg orbifold point ${\cal P}_{LG}$, and the
infinite volume limit $J=\infty$. Going to the zero-volume point,
for instance  along the path indicated by the arrows, corresponds
to a perfectly well-behaved string theory solution. Without
$\ap$-corrections, however, we would obtain a null singularity of
general relativity if we tried to go to zero-volume.}
{\epsfxsize2.2in\epsfbox{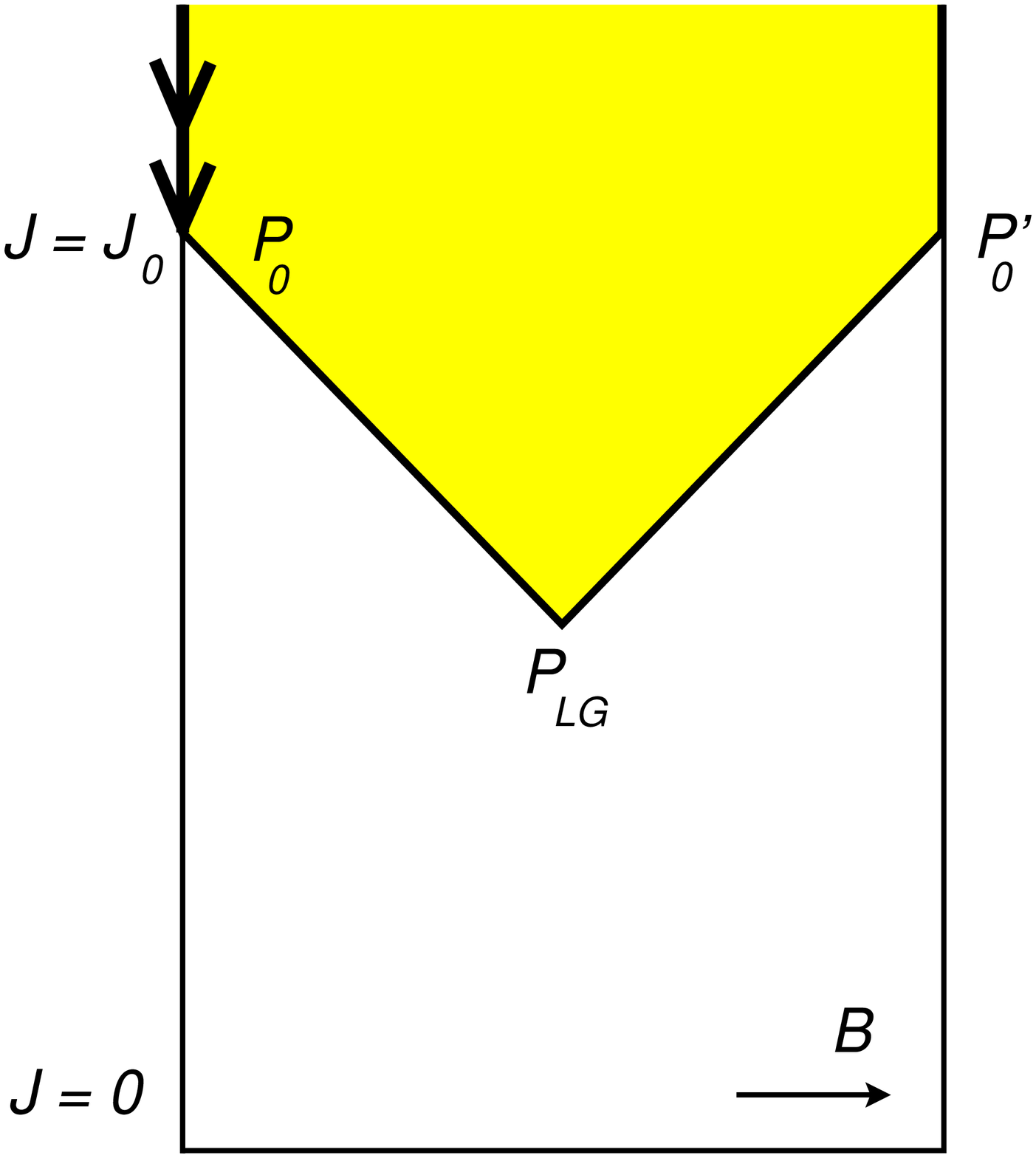}}

The vector multiplet moduli space and its metric have been
completely determined. Schematically, it is depicted in fig. 1.
Note that there are three interesting points: the infinite volume
limit, the Landau-Ginzburg orbifold point ({\it i.e. } the Gepner
point), and the zero-volume point ${\cal P}_0$, where D6-branes
wrapping the whole quintic become massless.

The metric is finite everywhere  except in the vicinity of ${\cal
P}_0$, where it has a logarithmic divergence caused by the light
D6-branes. Let us see more explicitly whether we can reach the
zero-volume point ${\cal P}_0$ in a finite light-cone time $\tilde
u$ without causing large curvatures in four dimensions, {\it i.e.}
with having $\tilde b$ in \planebrinkmaneinstein\ bounded by some
finite value.
The only non-trivial equation of motion is  \constraintbrinkman,
or in our case
\eqn\qconstraint{\tilde b =  4\pi {\cal G}_{AB}(\phi^C)\ \! \dot
\phi^A \ \! \dot \phi^B, \quad A, B, C=1,2. }
We can choose $\phi^1$ to be the `K\" ahler form' $J$ of the
quintic, and $\phi^2$ can be the `$B$-field period, $B$, as in
fig. 1. Let us also define $\rho = J-J_0$. If we choose $B$ to be
constant along the path in the moduli space, the equation
\qconstraint\ becomes
\eqn\qconstrainttwo{\tilde b = 4\pi {\cal G}_{11}\ \! \dot
\rho^2.}
Near the zero-volume point the metric has a logarithmic behavior,
and we may write
\eqn\qconstraintthree{\tilde b \  d\tilde u^2 \propto {\rm log} \
\! \rho \ d\rho^2.}
Because the integral of $({\rm log} \ \! \rho)^{1/2} $ from zero
to any finite positive $\rho$ is finite, we see that indeed, it is
possible to reach the zero-volume point within a  finite interval
of the light-cone time and  with the four-dimensional curvature
being small.

%%%%%%%%%%%%%%%%%%%%%%%%%%%%%%%%%%%%%%%%%%%%%%%%%%%%%%%%%%%%%%
\newsec{Conclusions}
%%%%%%%%%%%%%%%%%%%%%%%%%%%%%%%%%%%%%%%%%%%%%%%%%%%%%%%%%%%%%%
%gggggggggg

We have seen that understanding null singularities in string
theory does not always pose a much harder problem than
understanding static string compactifications. In particular, we
have seen that many null singularities have a perfectly
non-singular  description within the framework of
string theory at weak coupling.

\bigskip
\bigskip
\bigskip
\bigskip
\centerline{\bf Acknowledgements}
\medskip

We would like to thank Mina Aganagic, John M$\irc$Greevy, Sergei Gukov,
Veronika Hubeny, Shamit Kachru, Xiao Liu, Liam M$\irc$Allister, David
Morrison, Lubo\v s Motl, Nathan Seiberg,
 Steve Shenker, and Eva Silverstein for valuable discussions.
  M.F. is grateful to
the Harvard Theory Group and to the Harish-Chandra Research
Institute for hospitality.
The work of S.H. was supported by the DOE under
contract DE-AC03-76SF00515.
 The work of M.F. was supported by the DOE under
contract DE-AC03-76SF00515, by the A.P. Sloan Foundation, and by a
Stanford Graduate Fellowship.

%%%%%%%%%%%%%%%%%%%%%%%%%%%%%%%%%%%%%%%%%%%%%%%%%%%%%%%%%%%%
\appendix{A}{Setting Up The Coordinate System}
%%%%%%%%%%%%%%%%%%%%%%%%%%%%%%%%%%%%%%%%%%%%%%%%%%%%%%%%%%%%
%hhhhhhhhhh

In the appendix A, we will set up the coordinate system used in
section 3 and prove some of its important properties along the
way.

\subsec{The non-singular case}

In the following,  we will make two assumptions: (1) We will
assume that the spacetime is a connected manifold of Lorentzian
signature which admits a covariantly constant null vector. (Which
means that it is a pp-wave, a plane-fronted wave with parallel
rays.) This is true, in particular, in the case of
null-supersymmetric supergravity solutions without flux and with a
constant dilaton. (2) We will assume there are no closed causal
curves. The only reason we need this assumption is to make sure
that the null isometry $\CI$ defined below is non-compact. If we
wanted to accept also compact null isometries, we could relax this
condition.

Let us denote the covariantly constant null vector $l^\mu$. (In
principle, there might be more such vectors which would be
linearly independent, but we will use only one of them, denoted
$l^\mu$, in all of our considerations.) The covariantly constant
vector field $l^\mu$ may be used to define a scalar field $u$ at
any point $P$ in the spacetime as
\eqn\udefinition{u(P)=\int_{c(P, P_0)}  l_\mu \ {ds^\mu}, }
where $c(P, P_0)$ is a path connecting the point $P$ to some fixed
reference point $P_0$, and $ds^\mu$ is a line element along the
path. Because $l^\mu$ is covariantly constant, the integral does
not depend on the particular choice of $c(P, P_0)$, and the scalar
$u$ is well-defined.\foot{
Strictly speaking, this is true only if $\int_{\gamma}  l_\mu \
{ds^\mu}$ vanishes for all one-cycles $\gamma$ in the spacetime.
This will be the case in all examples we are interested in, and we
can simply assume that this requirement is satisfied.
Nevertheless, if $\int_{\gamma}  l_\mu \ {ds^\mu}$ does not vanish
for some one-cycle $\gamma$, it just means that $u$ is a
multi-valued scalar in the spacetime. (An example of such
spacetime would be a gravitational wave propagating in the
$S^1$-direction in, say, $\IR^{1,8}\times S^1$.) In this case, we
can always go to the covering space, where $u$ is single valued.
Since all the statements we make in this appendix and in section 3
are only for some restricted range of $u\in (u_{\rm a}, u_{\rm
b})$, it will not be important whether $u$ is single-valued or
multi-valued. Roughly speaking, this is because we can choose
$u_{\rm a}$ and $ u_{\rm b}$ such that the one-cycle $\gamma$ (for
which $\int_{\gamma}  l_\mu \ {ds^\mu}\ne 0$) intersects $u=u_{\rm
a}$
and $u=u_{\rm b}$. %
}  As a result, the spacetime will be foliated
by slices $\CS_u$ of constant $u$.

The hypersurfaces $\CS_u$ have the following property:
Any geodesic which is tangent to $\CS_u$ at
one point lies entirely inside that particular  $\CS_u$. This is a
simple consequence of the fact that the change of $u$ with the
affine parameter $\lambda$ of the geodesic can be written as
\eqn\udot{{du \over d\lambda} =  l_\mu \ {ds^{ \mu} \over
d\lambda^{~}} \ \!  .}
The scalar product of a covariantly constant vector with a tangent
vector of a geodesic does not change under parallel transport
along that geodesic. As a result, if $du/d\lambda$ vanishes at one
point, it will be zero at any other point of the geodesic.

The scalar $u$ is globally well-defined and we will use it as a
coordinate. In addition, we would like to define coordinates $v$
and $x^i$.

Start at an arbitrary surface $\CS_{u_0}$, corresponding to
$u=u_0$. The null Killing vector $l^\mu$ is, by definition
\udefinition, tangent to any surface $\CS_u$, and in particular,
it generates a  non-compact continuous  isometry $\CI$ which takes
$\CS_{u_0}$ to itself. (It is non-compact because we have assumed
that there are no closed causal curves.) Since $l^\mu$ is
non-trivial and covariantly constant, it is everywhere
non-vanishing. This means that the isometry $\CI$ acts freely. As
a result, there exists a smooth cross-section $\Sigma_{u_0}$ of
$\CS_{u_0}$ such that (1) no point on $\Sigma_{u_0}$ is the image
of any other point of $\Sigma_{u_0}$ by a non-trivial isometry
action $\CI$, and (2) $\Sigma_{u_0}$ together with its images by
$\CI$ covers the whole $\CS_{u_0}$. (Clearly, $\Sigma_{u_0}$ will
be homeomorphic to the coset space $\CS_{u_0}\ \! / \ \! I$.)

\ifig\quintic{Setting up a coordinate system is not an easy task
if your tools are curved.} {\epsfxsize3.5in\epsfbox{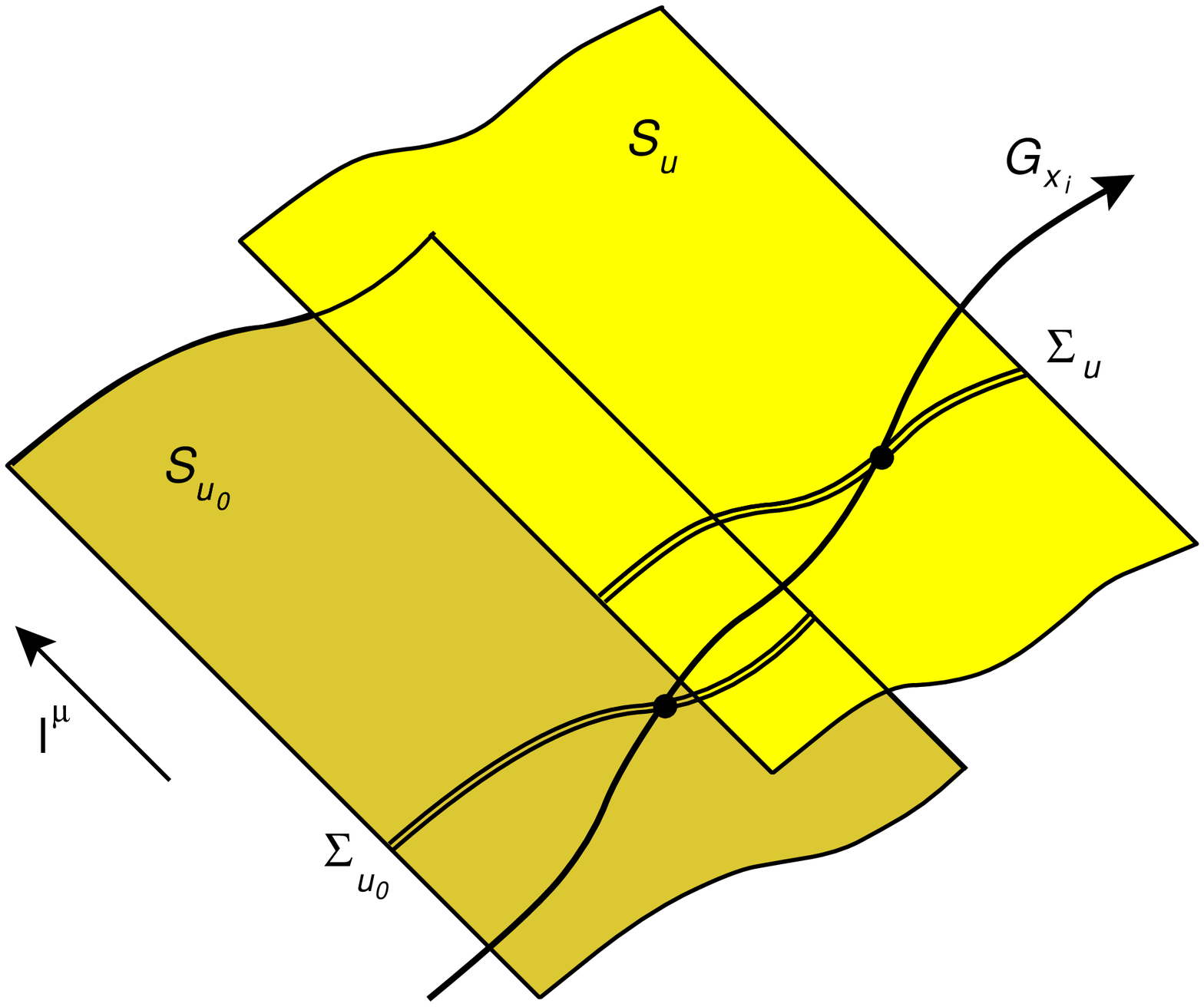}}

We can set up an arbitrary coordinate system on $\Sigma_{u_0}$
with coordinates $x^i$.\foot{
%%%%
If $\Sigma_{u_0}$ is topologically non-trivial, we might need to
use more patches of coordinates $x^{i}_{(a)}$, labelled by an
index $a$. We will not express this fact explicitly, in order not
to obscure the notation even more.
%%%%
} Now, we will try to extend these coordinates also to some
cross-sections $\Sigma_{u}$ of other hypersurfaces $\CS_u$. At
every point of $\Sigma_{u_0}$, we will find a null direction
normal to $\Sigma_{u_0}$ and independent of $l^\mu$. These
directions define a congruence of null geodesics $\CG_{x^i}$
%originating from $\Sigma_{u_0}$,
which will necessarily intersect
all the surfaces $\CS_u$. We  define the coordinates $x^i$ along
any geodesic to be equal to the value of $x^i$ at the intersection
of the geodesic with $\Sigma_{u_0}$. This definition will be
sensible only in some range ($u_{\rm a}, u_{\rm b}$) of $u$,
$u_{\rm a} < u_0 < u_{\rm b}$, because the null congruences will
almost inevitably have some caustics. The fact that $\Sigma_{u_0}$
is smooth guaranties that $u_{\rm a} < u_0$ and $u_0 < u_{\rm b}$.
>From now on we will restrict our attention to the range ($u_{\rm
a}, u_{\rm b}$).

So far we have defined $u$ globally and $x^i$ on one cross-section
$\Sigma_{u}$ of every $\CS_{u}$, $u \in (u_{\rm a}, u_{\rm b}) $.
Now we will extend the definition of $x^i$ to any spacetime point
with $u \in (u_{\rm a}, u_{\rm b}) $ in a simple way manner using
the null isometry $\CI$ generated by $l^\mu$. At every point of
every surface $\Sigma_{u}$, $u \in (u_{\rm a}, u_{\rm b}) $, we
construct a null geodesic $\tilde \CG_{u, x^i}$  in the
$l^\mu$-direction, and define $x^i$ to be constant along each of
these geodesics. We will use the same geodesics also in the
following step.

Having specified $u$ and $x^i$ for every point of interest, {\it
i.e.} at every point with $u \in (u_{\rm a}, u_{\rm b})$, the only
other coordinate to be defined is $v$. We can choose $v$ to be
equal to zero at all the cross-sections $\Sigma_{u}$ constructed
previously. Clearly, the  direction in which only $v$ will
increase is along the geodesics $\tilde \CG_{u, x^i} $. Let us
calibrate the affine parameter $\lambda_{u, x^i}$ of each $\tilde
\CG_{u, x^i} $ in such a way that the tangent vector $ds^\mu /
d\lambda_{u, x^i}$ equals $l^\mu$ at any $\Sigma_{u}$. Now define
$v$ along any geodesic $\tilde \CG_{u, x^i}$ to be equal to the
corresponding value of the affine parameter $\lambda_{u, x^i}$.
This guarantees that the isometries $\CI$ generated by $l^\mu$
will be realized as constant shifts of $v$ without changing $u$
and $x^i$. In other words, $dv$ will be a  Killing vector.

In the coordinate system we have just constructed, there are
important simplifications in the metric. If we write its general
form as
\eqn\generalmetric{ds^2 = g_{uu}\ \! du^2 + g_{vv}\ \! dv^2 + 2 \
\!  g_{uv}\ \! du \ \! dv + 2\ \! g_{ui}\ \!  du\ \!  dx^i + 2\ \!
g_{vi}\ \! dv \ \!  dx^i + g_{ij} \ \! dx^i\ \! dx^j,  }
we notice the following properties:

$\bullet \ \  $ None of the metric coefficients depends on $v$,
because $\del_v$ is by construction a Killing vector.

$\bullet \ \  $ The coefficient $g_{uu}$ vanishes, because
$\del_u$ is a vector tangent to some null geodesic which is an
$\CI$-translation of one of the  null geodesics $\CG_{x^i}$.

$\bullet \ \  $ The coefficient $g_{vv}$ vanishes, because
$\del_v$ is a vector tangent to some null geodesic $\tilde \CG_{u,
x^i}$.

$\bullet \ \  $ The coefficient $g_{uv}$ is equal to $-1$, by the
definition of $u$ \udefinition\ and the definition of $v$.

$\bullet \ \  $ The coefficient $g_{vi}$ vanishes, because any
vector (in our case $\del_{x^i}$) tangent to a surface of constant
$u$ is perpendicular to $\del_v \sim l^\mu$.

$\bullet \ \  $ The last statement will show now is that $g_{ui}$
vanishes. The geodesic equation for $\CG_{x^i}$ can be written as
\eqn\geodesic{g_{\mu \nu}  {d^2 s^\nu\over d\lambda^2} + g_{\mu
\nu} \Gamma^{\nu}_{\rho \sigma} \ \! {ds^\rho \over d\lambda} \ \!
{ds^\sigma \over d\lambda} = 0.}
The only coordinate that varies along $\CG_{x^i}$ is $u$, so the
equation becomes
\eqn\geodesictwo{g_{\mu u}  {d^2 u \over  d\lambda^2} + g_{\mu
\nu} \Gamma^{\nu}_{uu} \ \! \left({du \over d\lambda}\right)^2  =
0.}
Because $g_{uu}$ vanishes, the Christoffel symbol simplifies, and
we get
\eqn\geodesicthree{g_{\mu u}  {d^2 u \over  d\lambda^2} + g_{\mu
u, u}  \ \! \left({du \over d\lambda}\right)^2 = 0.}
To determine $d^2 u / d\lambda^2$, we can set $\mu = v$ in
\geodesicthree,
\eqn\geodesicfour{g_{v u}  {d^2 u \over  d\lambda^2} + g_{v u, u}
\ \! \left({du \over d\lambda}\right)^2 = 0.}
Since $g_{v u} $ is a constant,  $d^2 u / d\lambda^2=0$ must
vanish. This means that $u$ is a good affine parameter for the
geodesic $\CG_{x^i}$, as could have been anticipated. Returning
back to \geodesicthree, we see that
\eqn\geodesicfive{g_{\mu u, u} = 0}
for any $\mu$. In particular, we can take $\mu = i$.

The geodesic $\CG_{x^i}$ was constructed in such a way that at
$u=u_0$ it is perpendicular to $\Sigma_{u_0}$, which means that
 $g_{i u}$ vanishes at $\Sigma_{u_0}$. Equation \geodesicfive\ then
implies that $g_{i u}$ vanishes at any $\Sigma_{u}$. It is now
trivial to extend this result to the whole spacetime between
$u=u_{\rm a}$ and $u=u_{\rm b}$, since the coordinates $x^i$ have
been defined by $\CI$-translations of the coordinates $x^i$  at
various $\Sigma_{u}$.

We have just shown that in our coordinate system, the metric in
the region between  $u = u_{\rm a}$ and $u =  u_{\rm b}$ takes the
form
\eqn\ppmetriccopy{ds^2 = -2\ \! du\ \! dv + h_{ij}(u,x^k)\ \!
dx^i\ \! dx^j,}
which was the goal of this section.

\subsec{The singular cases}

Even if the spacetime is singular, having a covariantly constant
vector implies that we can define $u$ in the same way as in the
previous section. This means that the spacetime will still be
foliated by surfaces $\CS_u$ of constant $u$. Now there are two
possibilities.

(1) If there exists a family of non-singular $\CS_u$ which
degenerates at some $u_s$, then we can apply the results of the
previous section to the non-singular region, and generally, we can
study the properties of this singularity by looking at the path in
the space of solutions of 8d Euclidean gravity in the spirit of
section 4. There are however two pathological cases which cannot
be understood in this way. One of them is an orbifold singularity
which corresponds to a $\IZ_2$ action reflecting the
$u$-direction, and which introduces closed causal curves. The
other one is the case where there are infinitely many conjugate
points near the singularity, which implies that in no open
interval $(u_0, u_s)$ touching the singularity we can use one set
of coordinates leading to \ppmetriccopy\ everywhere.
Heuristically, this corresponds to a gravitational wave with an
unbounded frequency.

(2) Even if every $\CS_u$ is singular, there should be some family
of $S_u$ where the singular loci are at least codimension one in
$S_u$, since otherwise we would not even know how to define the
spacetime. If is quite possible that there is a suitable
generalization of the arguments from the previous section which
can be applied to this case as well. This would go, however,
beyond the scope of the present paper.

%In this region the discussion from the previous section can be
%applied as well, with the modification that the cross-sections
%$\Sigma_u$ will have some singularities (for example orbifold
%singularities). If we are willing to accept that, there will be
%not much difference from the case (1).

%%%%%%%%%%%%%%%%%%%%%%%%%%%%%%%%%%%%%%%%%%%%%%%%%%%%%%%%%%%%
\appendix{B}{Stability}
%%%%%%%%%%%%%%%%%%%%%%%%%%%%%%%%%%%%%%%%%%%%%%%%%%%%%%%%%%%%

From a certain point of view, the stringy resolutions of null
singularities discussed in section 7 are just plane waves with
some number of scalars varying in the same light-like directions.
For this reason, it is obvious that they are stable (for a recent
discussion of the stability of plane waves see \refs{\brecher, \MarolfBX}. )

It is, however,  quite interesting to see intuitively why effects
considered in a great detail in \HorowitzMW\ do not pose a problem
here. In the case of shrinking Calabi-Yau manifolds, we cannot use
any arguments based on ten-dimensional supergravity when the size
of the Calabi-Yau becomes of order the string length. However, we
can still ask whether there is any instability related to the
evolution of the spacetime before the Calabi-Yau shrinks to a
string size.

The closest simple analog of such spacetimes would be an
$S^1$-fibration over a nine-dimensional plane wave, which for $u
\equiv {\rm x^+}$ smaller than some $x^+_c$ looks exactly as the
parabolic orbifold, but where the after $x^+_c$ the circle stopes
shrinking and expands again. Such spacetimes have been constructed
in \FabingerKR.

\ifig\stability{(a) A schematic picture of the parabolic orbifold,
showing only coordinates $\rmx^+$ and $\rmx^-$.
 Close to the singularity, images of any particle
become infinitely dense. (b) If one cuts off the spacetime at some
finite $x^+_c < 0 $ and replaces it with a plane wave where the
circle expands again, the images never come too close to each
other, and the resulting spacetime is stable. The part of the
geometry with ${\rm x}^+ > {\rm x}^+_c $, not shown in this
figure, has a non-zero curvature. }
{\epsfxsize5in\epsfbox{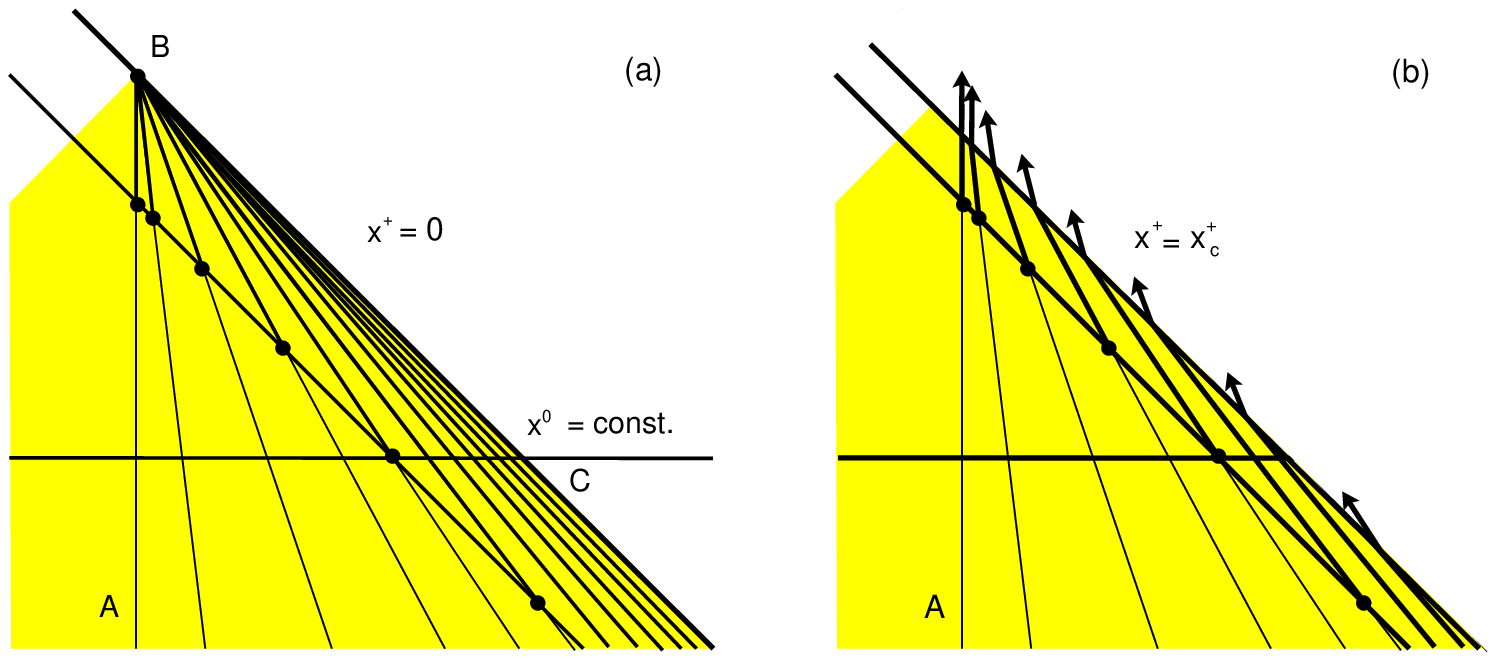}}

To see intuitively why these spacetimes are stable but the
parabolic orbifold is not, consider the situation in fig. 3. If
there is a massive particle in the parabolic orbifold (a) which
starts at some spacetime point A and reaches the singularity at
point B, then there will be also infinitely many images of the
same particle which are all in the past light cone of B, and which
reach the singularity at point B. At any slice of constant time
${\rm x^0}=$ const., there will be a finite density of images
everywhere  except close to the singularity at a point denoted C.
It is precisely the infinite concentration of images at C which
causes a large backreaction, and as a consequence, an instability
of the singularity itself.

If one cuts off the singularity (b) at some light-cone-time $x^+_c
< 0 $ and replaces it with a plane wave where the circle expands
again, as in \FabingerKR, the image particles never come too close
to each other, and the particle density is finite everywhere. This
is in accord with the usual intuition that a compactification on a
finite-size circle should be stable even when the size of the
circle varies with time. A spacetime of this type would be stable
even if we replaced the $S^1$ with, say $T^6$. There is no need of
a large number of non-compact directions, unlike in the case of
the null-brane considered in \refs{ \LiuKB  \FabingerKR -
\HorowitzMW}. This can be seen by a simple analysis of the
Kaluza-Klein modes in this kind of geometry. Kaluza-Klein
excitations will always have finite energy and finite energy
density, and provided that the string coupling constant is not too
large, their scattering can be studied perturbatively. Of course,
there will be also scattering processes which produce a
finite-size black holes. This is however the same situation as in
flat space, and it cannot be considered to be an instability of
the spacetime itself.

\listrefs

\bye